\documentclass[a4paper,fleqn,usenatbib]{mnras}
\usepackage{newtxtext,newtxmath}
\usepackage[T1]{fontenc}
\usepackage{ae,aecompl}

\usepackage{mn2e-breakabs}
\usepackage{graphicx}
\usepackage{subfigure}
\usepackage{times}
\usepackage{caption}
\usepackage{rotating}
\usepackage{rotfloat}
\usepackage{multicol}
\usepackage{multirow}
\usepackage{array}
\usepackage{booktabs}
\usepackage{color}
\usepackage{afterpage}
\usepackage[multidot]{grffile}
\usepackage{bm}
\usepackage{float}
\usepackage{hyperref}
\providecommand{\adsurl}[1]{\href{#1}{ADS}}

\newcommand{\Hunit}{\,{\rm km}\,{\rm s}^{-1}\,{\rm Mpc}^{-1}}

\def\la{\mathrel{\mathpalette\fun <}}
\def\ga{\mathrel{\mathpalette\fun >}}
\def\fun#1#2{\lower3.6pt\vbox{\baselineskip0pt\lineskip.9pt
        \ialign{$\mathsurround=0pt#1\hfill##\hfil$\crcr#2\crcr\sim\crcr}}}

\newcommand{\be}{\begin{equation}}
\newcommand{\ee}{\end{equation}}
\newcommand{\ba}{\begin{eqnarray}}
\newcommand{\ea}{\end{eqnarray}}
\newcommand{\simgt}{\,\hbox{\lower0.6ex\hbox{$\sim$}\llap{\raise0.6ex\hbox{$>$}}}\,}
\newcommand{\simlt}{\,\hbox{\lower0.6ex\hbox{$\sim$}\llap{\raise0.6ex\hbox{$<$}}}\,}

\begin{document}

\title[Robustness of the covariance matrix]
{
Robustness of the covariance matrix for galaxy clustering measurements
}

\author[Baumgarten \& Chuang]{
  \parbox{\textwidth}{
  Falk Baumgarten$^{1,2}$,
  Chia-Hsun Chuang$^{3,1}$\thanks{E-mail: chuangch@stanford.edu}
}
  \vspace*{4pt} \\
  $^1$ Leibniz-Institut f\"ur Astrophysik Potsdam (AIP), An der Sternwarte 16, D-14482 Potsdam, Germany\\
  $^2$ Humboldt-Universit\"at zu Berlin, Institut f\"ur Physik, Newtonstrasse 15, D-12589 Berlin, Germany\\
  $^3$ Kavli Institute for Particle Astrophysics and Cosmology \& Physics Department, Stanford University, Stanford, CA 94305, USA\\
}

\date{\today} 

\maketitle

\begin{abstract}
We present a study on the robustness of the covariance matrix estimation for galaxy clustering measurements depending on the cosmological parameters and galaxy bias. 
To this end, we have produced 9000 galaxy mock catalogues relying on the effective Zel’dovich approximation implemented in the EZmocks computer code, using different input cosmological models and bias parameters.
The reference catalogue has also been produced with this code making our study insensitive to the approximation at least on a relative-qualitative level. 
Our findings indicate that the covariance matrix is insensitive to the input power spectrum (including $\sigma_8$), as long as the 2- and 3-point galaxy clustering measurements agree with the given data.
In fact, the covariance matrix shows a bias at small scales ($r\lesssim40 h^{-1}$Mpc) when the chosen galaxy bias parameters yield a 3-point statistics, which is not compatible with the reference one within the error bars, even though the 2-point statistics agrees within 1\%.
Nevertheless, the error becomes negligible at large scales making the covariance matrix still reliable for data analysis using only measurements in that regime (e.g., measuring baryon acoustic oscillations). 

High precision in cosmological parameter estimation is expected for covariance matrices extracted from mock galaxy catalogues which take accurately into account both the 2- and the 3- point statistics. This is independent on whether this is achieved by using the right cosmology and galaxy bias (which are not a priori known) or just any combination of both fitting the net observed galaxy clustering.
\end{abstract}


\section{Introduction} \label{sec:intro}

The scope of galaxy redshift 
surveys has dramatically increased in the last two decades. The astrophysical community has been extracting cosmological information from completed surveys, such as the 2dF Galaxy Redshift Survey\footnote{http://www2.aao.gov.au/2dfgrs/} (2dFGRS) \citep{Colless:2001gk,Colless:2003wz}, the Sloan Digital Sky Survey\footnote{http://www.sdss.org} (SDSS, \citealt{York:2000gk,Eisenstein:2011sa}), and WiggleZ\footnote{ http://wigglez.swin.edu.au/site/} \citep{Drinkwater:2009sd, Parkinson:2012vd}, as well as on-going surveys, e.g. DES\footnote{http://www.darkenergysurvey.org} (Dark Energy Survey) and eBOSS\footnote{http://www.sdss.org/sdss-surveys/eboss/} (Extended Baryon Oscillation Spectroscopic Survey).
There are also new upcoming ground-based and space experiments, such as
4MOST\footnote{http://www.4most.eu/} (4-metre Multi-Object Spectroscopic Telescope, \citealt{deJong:2012nj}), DESI\footnote{http://desi.lbl.gov/} (Dark Energy Spectroscopic Instrument,\citealt{Schlegel:2011zz,Levi:2013gra}),
HETDEX\footnote{http://hetdex.org} (Hobby-Eberly Telescope Dark Energy Experiment, \citealt{Hill:2008mv}),
J-PAS\footnote{http://j-pas.org} (Javalambre Physics of accelerating universe Astrophysical Survey, \citealt{Benitez:2014ibt}), 
LSST\footnote{http://www.lsst.org/lsst/} (Large Synoptic Survey Telescope, \citealt{Abell:2009aa}), 
Euclid\footnote{http://www.euclid-ec.org } \citep{Laureijs:2011gra}, 
and WFIRST\footnote{http://wfirst.gsfc.nasa.gov} (Wide-Field Infrared Survey Telescope, \citealt{Spergel:2013tha}).

Covariance matrices are essential for analysing the clustering signal drawn from these surveys. 
Using mock galaxy catalogues is considered the most reliable way to estimating covariance matrices. 
A straightforward approach to create mock catalogues is running $N$-body cosmological simulations.
However, the total run-time and memory required to generate a large suit of simulations make this effort prohibitive in most of the cases, hence their use for ongoing and future surveys is impractical. 
Recent techniques permit to speed up $N$-body codes (see COLA\footnote{COLA (COmoving Lagrangian Acceleration simulation)} \citealt{Tassev:2013pn}, FastPM \citealt{Feng:2016yqz}, and PPM-GLAM \citealt{Klypin:2017iwu}). However, the memory requirements are still large with these methods.
Alternatively, mock catalogues can be produced by approximative methods, e.g., log-normal mock catalogues \citep{Coles:1991if}, Peak-Patch \citep{Bond:1993we}, PTHalos \citep{Scoccimarro:2001cj,Manera:2012sc,Manera:2014cpa}, PINOCCHIO\footnote{PINOCCHIO (PINpointing Orbit-Crossing Collapsed Hierarchical Objects)} \citep{Monaco:2001jg,Monaco:2013qta}, PATCHY\footnote{PATCHY (PerturbAtion Theory Catalog generator of Halo and galaxY distributions)} \citep{Kitaura:2013cwa,Kitaura:2014mja,Vakili:2017rsp}, QPM\footnote{QPM (Quick Particle Mesh)} \citep{White:2013psd}, Halogen: \citep{Avila:2014nia}, and EZmock\footnote{EZmock (Effective Zel'dovich approximation mock catalogues)} \citep{Chuang:2014vfa}. Some review and comparisons of different methodologies can be found in \cite{Chuang:2014toa}.

The bias of the precision matrix (inverted covariance matrix) due to the finite number of mock catalogues used to construct the covariance matrix has been evaluated in a number of studies, e.g. see \cite{Hartlap:2006kj,Taylor:2012kz,Dodelson:2013uaa,Percival:2013sga,Taylor:2014ota,Blot:2014pga,Blot:2015cvj}. 
The methodologies to smooth, improve, or analytically model the covariance matrix constructed by a smaller number of mock catalogues have been developed, e.g. see \cite{Pope:2007vz,Chuang:2011fy,Mohammed:2014lja,Paz:2015kwa,Padmanabhan:2015vlf,OConnell:2015src,Pearson:2015gca}. 
Some progresses in the direction of rescaling the covariance matrix constructed with smaller volume simulations have been made, see \cite{Cole:1996hb,Schneider:2011wf,Takada:2013bfn,Howlett:2017vwp}.
Model-dependent covariance matrices have been studied by various literature, e.g. see \cite{Eifler:2008gx,Labatie:2012ue,Morrison:2013tqa,White:2015sya}. However, the effect would not be critical when the variance of data is small due to huge survey volume.

In this work, we want to understand the impact of the accuracy of the mock catalogues on the robustness of the covariance matrix.
We examine separately the impacts of varying two factors: firstly, the power spectrum of
initial conditions which deviate from the true cosmology and secondly, the usage of mock galaxy catalogues whose three-point clustering statistics are not reproducing the observed ones.
In principle, the covariance matrix can be predicted by
\begin{equation}
C_{ij}=\frac{2P^2(k_i)}{N_{k_i}}\delta_{ij}+T(k_i,k_j),
\end{equation}
where $N_{k_i}$ is the number of modes in the $k$ bin, $\delta_{ij}$ is Kronecker delta, and $T$ is the bin-averaged trispectrum (e.g., see \citealt{Bernardeau:2001qr} for a detail review). While it is not easy to adjust the trispectrum in practice, we adjust the bispectrum and observe the impacts on the covariance matrices.

We use the EZmock methodology \citep{Chuang:2014vfa} to generate mock galaxy catalogues for two reasons.
First, the computation of the EZmocks demands a minimum run-time, i.e., three fast Fourier transform (FFT) to compute the displacement field in three directions and populate galaxies nearby each grid point with some random assignment process, and a minimum memory requirement, i.e., a few arrays of the same size as the grid used by FFT to store the information of the displacement field and the number of generated galaxies. Secondly, the flexibility of the effective bias model for adjusting galaxy clustering statistics is critical for this study. By having an agreement in the clustering measurements among galaxy catalogues with different conditions, e.g., different input power spectrum, we are able to compare the covariance matrices self-consistently.

If the ongoing and upcoming large galaxy surveys will demand massive production of mock catalogues with huge volume, simple and efficient, but accurate methods will be favoured to cover those needs. To this end, we want to understand the requirements of constructing mock catalogues by examining the robustness of the covariance matrix estimate. Our study should help in designing a strategy for using limited resources to analyse the large-scale structure survey data.

This paper is organised as follows. In Section \ref{sec:ezmock}, we describe the EZmock simulations generated for this study.  
In Section \ref{sec:results}, we show the comparisons of covariance matrices constructed by different sets of EZmock simulations.
We summarise and conclude in Section \ref{sec:conclusion}.

\section{EZmock simulations}
\label{sec:ezmock}

\subsection{Description of EZmocks} 
\label{sec:desciption}

In this section, we describe the methodology to construct EZmock galaxy catalogues, which is slightly different from the original method \citep{Chuang:2014vfa}. 
It is based on the dark matter density field on a grid using the Zel'dovich approximation.
A particle located at Lagrangian position $\boldsymbol q$ will be mapped to its Eulerian position $\boldsymbol x$ at cosmic time $t$ by the displacement field $\boldsymbol\Psi(\boldsymbol q,t)$, i.e.,
\begin{equation}
\boldsymbol x(\boldsymbol q,t)=\boldsymbol q+\boldsymbol\Psi(\boldsymbol q,t).
\end{equation}
The first order Lagrangian perturbation theory solution to the equations of motions is given by the Zel'dovich approximation (for a review, see, e.g., \citealt{Bernardeau:2001qr}).
The displacement field in the ZA is given by
\begin{equation}
\boldsymbol\Psi(\boldsymbol q)=\int{\frac{d^3k}{(2\pi)^3}e^{i\boldsymbol k\cdot \boldsymbol q}\frac{i\boldsymbol k}{k^2}\hat\delta(\boldsymbol k)},
\end{equation}
where $\hat\delta(\boldsymbol k)$ is the fractional density perturbation in Fourier-space.
We construct the displacement field using the ZA to the redshift of given halo/galaxy sample, i.e., $z = 0.5618$ in this study. 
We use two parameters to describe the probability distribution function (PDF) of the expected output catalogue. One parameter, \texttt{n\_density}, determines the number density of the output catalogue and the other parameter, \texttt{pdf\_slope}, determines the slope of the PDF. We model the PDF by
\begin{equation}
P(n) = B A^n,
\end{equation}
where $n$ is the number of objects in a cell, $A\equiv$ \texttt{pdf\_slope}, and $B$ is the normalization constant to obtain the desired \texttt{n\_density}.
We then perform a PDF mapping procedure between the ZA density field and the expected output catalogue. 
The density field is obtained using the cloud-in-cells particle assignment scheme (CIC, e.g., \citealt{Hockney1981}).
We introduce some scatter before the mapping procedure in a way which is different from the original EZmock paper. The new scattering formula is  
\begin{equation}
\rho_{s}(\boldsymbol r) = 
\left\{ \begin{array}{ll}
         (1-\exp(-\rho_{o}(\boldsymbol r)/\rho_a)(1+G(\lambda)) & \mbox{if $G(\lambda) \ge 0$};\\
        (1-\exp(-\rho_{o}(\boldsymbol r)/\rho_a)\exp(G(\lambda)) & \mbox{if $G(\lambda) < 0$},
\end{array} \right.
\end{equation}
        
where $\rho_s(\boldsymbol r)$ and $\rho_o(\boldsymbol r)$ are the ZA density field after and before the scattering, respectively. $G(\lambda)$ is a random number drawn from the Gaussian distribution with width $\lambda$. The exponential function, $\exp(G(\lambda))$, is used to avoid negative densities. The term including $\rho_a$ introduces a nonlinear mapping which converges to 1 when $\rho_{o}(\boldsymbol r)$ is large. It plays a similar role as the density saturation parameter in the original EZmock paper.
Furthurmore, we apply a density cut (called ``density threshold'' in \citealt{Chuang:2014vfa}) on ZA density field before applying the scatter formula to make sure that there is no object produced in the low density region.
The mock catalogue is then populated following a CIC distribution (see \citealt{Chuang:2014vfa} for details).
Finally, we assign the velocity by multiplying the displacement field by a factor, \texttt{vel\_ratio}, and combining it with a 3-D gaussian random motion with width \texttt{vel\_random}.
\texttt{vel\_ratio}  is computed by
\begin{equation}
\texttt{vel\_ratio} = f(z)H(z)/(1+z), 
\end{equation} 
where $f(z)$ and $H(z)$ are the growth rate and Hubble parameter, respectively, at the redshift $z$ of the simulation box.
Thus, we have 5 effective bias parameters (\texttt{n\_density}, \texttt{pdf\_slope}, $\lambda$, \texttt{density\_cut}, and \texttt{vel\_random}) to adjust EZmocks.

In contrast to \cite{Chuang:2014vfa}, we do not modify the input power spectrum for calibration since we want to observe the impact of different cosmological models, as explained in the next section.

\subsection{Simulation Setup}
\label{sec:setup}
We use the same input power spectrum as the one used by \cite{Chuang:2014vfa}, i.e. $\Lambda$CDM Planck cosmology with \{$\Omega_{\rm M}=0.307115, \Omega_{\rm b}=0.048206,\sigma_8=0.8288,n_s=0.96$\}, and a Hubble constant ($H_0=100\,h\Hunit$) given by  $h=0.6777$. The output galaxy catalogues are at redshift $z=0.5618$.
Again, following \cite{Chuang:2014vfa}, we choose the number density $3.5\times 10^{-4}$ $h^3\,{\rm Mpc}^{-3}$, which is similar to that of the BOSS galaxy sample at $z\sim0.5$. The simulation boxes are 2.5$h^{-1}$Gpc each side. The particle mesh used is $960^3$.

We construct three sets of 3000 EZmock simulation boxes with three sets of parameters. Table \ref{table:parameters} lists the values of these parameters. The first set is the reference, of which the clustering statistics are similar to those in \cite{Chuang:2014vfa}, but they are not the same since the EZmock methodology we are using here is slightly different, as mentioned above.

\begin{table}
 \begin{center}
  \begin{tabular}{ |c|c|c|c| }    
  \hline
 parameter & set 1 & set 2 & set 3 \\ \hline
   $\sigma_8$               & 0.8225    & 0.7403   & 0.8225    \\ \hline
    \texttt{n\_density}    & 3.5E-4    & 3.5E-4    & 3.5E-4     \\ \hline
  \texttt{pdf\_slope} & 0.130    & 0.115   & 0.144     \\ \hline
  $\lambda$    & 2.8500     & 2.8564  & 0.0010     \\ \hline
  \texttt{density\_cut}   & 0.0      & 2.1963  & 2.9300    \\ \hline
  \texttt{vel\_random}     & 0.0      & 63.2    & 0.0      \\ \hline
   \end{tabular}
 \end{center}
   \caption{The three sets of EZmock parameters used in this study. We construct 3000 EZmock simulation boxes for each set of parameters (i.e. 9000 boxes in total).}
  \label{table:parameters}
 \end{table}

\begin{figure}
\begin{center}
 \subfigure{\includegraphics[width=0.49 \textwidth]{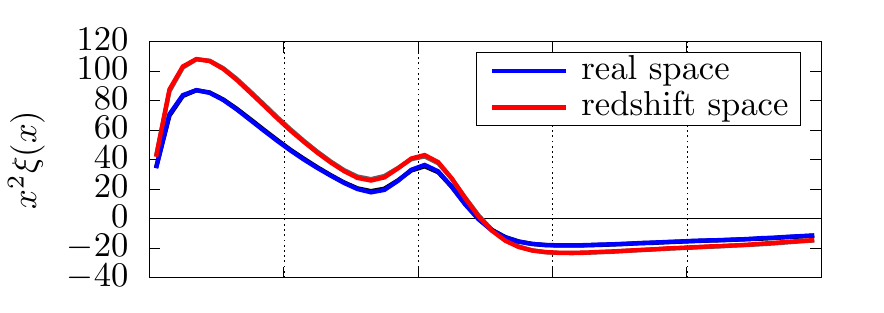}}
 \subfigure{\includegraphics[width=0.49 \textwidth]{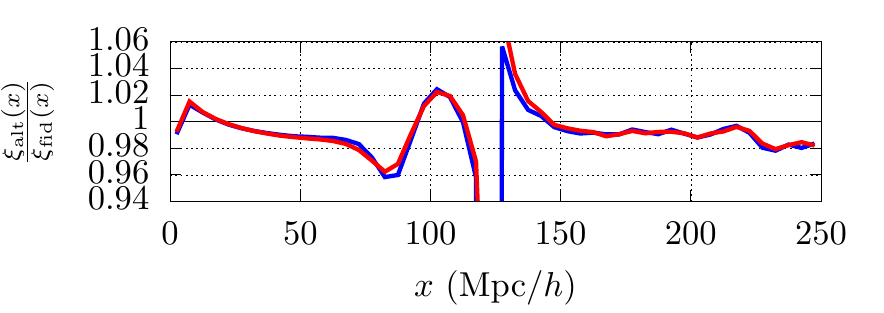}}
\end{center}
\caption{The upper panel: comparison between the mean correlation function of the set of 3000 EZmock boxes with 10\% smaller $\sigma_8$ and the one of the reference set. We show comparisons in both real space and redshift space. The references are shown with black and grey lines, but they are hardly visible since the second set has almost perfect agreement with the reference. The bottom panel: the ratios of the means of these two sets of EZmocks. From now on, we label the reference one with ``fid'' and the other one with ``alt''.}
\label{fig:cf_s8}
\end{figure}

\begin{figure}
\begin{center}
 \subfigure{\includegraphics[width=0.49 \textwidth]{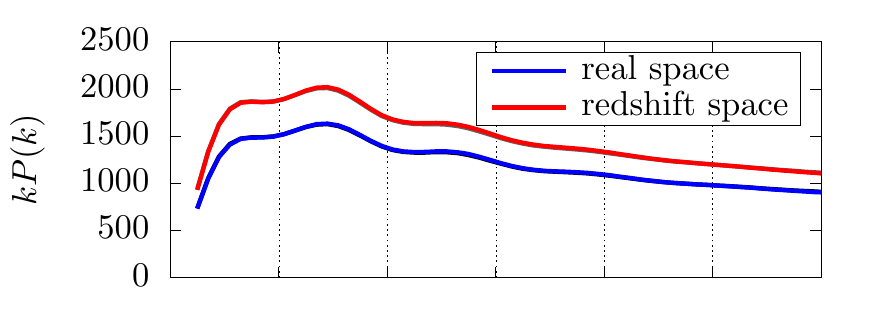}}
 \subfigure{\includegraphics[width=0.49 \textwidth]{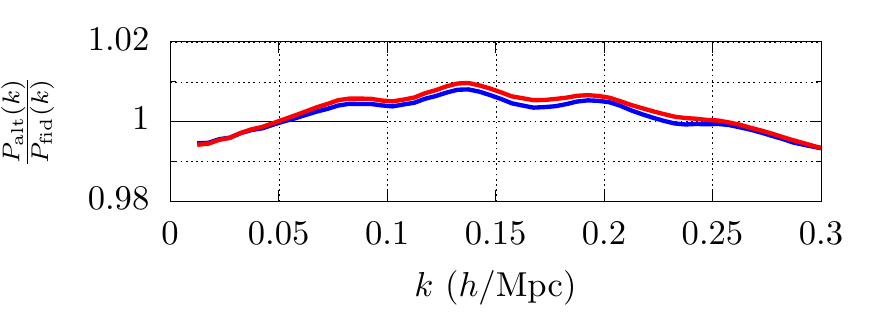}}
\end{center}
\caption{The upper panel: comparison between the power spectrum of the set of 3000 EZmock boxes with 10\% smaller $\sigma_8$ and the one of the reference set. We show comparisons in both real space and redshift space. The references are shown with black and grey lines, but they are hardly visible since the second set has almost perfect agreement with the reference. The bottom panel: the ratios of the means of these two sets of EZmocks. We label the reference one with ``fid'' and the other one with ``alt''.}
\label{fig:pk_s8}
\end{figure}

\begin{figure}
\begin{center}
 \subfigure{\includegraphics[width=0.49 \textwidth]{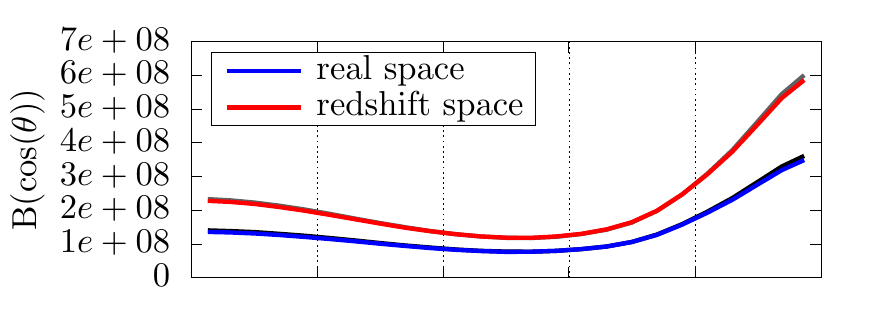}}
 \subfigure{\includegraphics[width=0.49 \textwidth]{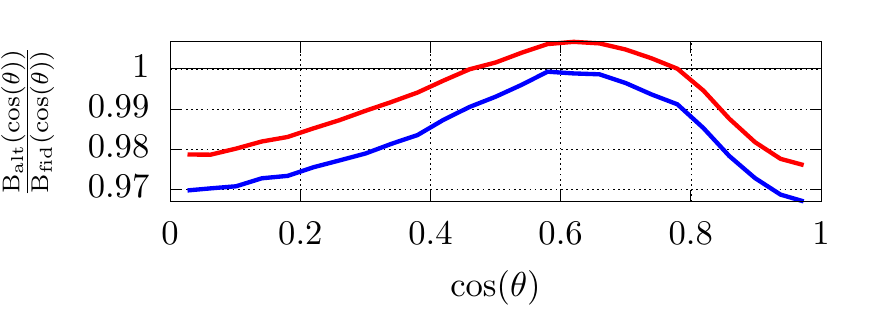}}
\end{center}
\caption{The upper panel: comparison between the bispectrum of the set of 3000 EZmock boxes with 10\% smaller $\sigma_8$ and the one of the reference set. The configuration of the bispectrum is $\{k_1=0.1,k_2=0.2\}$. We show comparisons in both real space and redshift space. The references are shown with black and grey lines, but they are hardly visible since the second set has almost perfect agreement with the reference. The bottom panel: the ratios of the means of these two sets of EZmocks. We label the reference one with ``fid'' and the other one with ``alt''.}
\label{fig:bk_s8}
\end{figure}

The second set of 3000 EZmocks has a different amplitude in the input power spectrum, i.e. 10\% smaller $\sigma_8$ than the reference one. We choose the EZmock bias parameters of the second set to have the same mean 2-point and 3-point clustering statistics as the reference set of EZmocks. Fig.\,\ref{fig:cf_s8}, \ref{fig:pk_s8}, and \ref{fig:bk_s8} show the comparison of the correlation function, power spectrum, and bispectrum of these two sets of mocks (the reference set and the set with smaller $\sigma_8$). One can see that their agreement is within 1\% of the power spectrum for $k<0.3$; within $2\%$ in most scales of the 2-point correlation function; within 3\% of the bispectrum. The fittings cannot be perfect because a different $\sigma_8$ results in slightly different shape of the clustering. A smaller $\sigma_8$ results in weaker damping of baryon acoustic oscillations (BAO), so that one can still see the BAO features in the ratio plots of the correlation functions and power spectra. For the bispectrum, a smaller $\sigma_8$ produces a less curved ``U'' shape, as can easily be observed in the plot of ratios. 

\begin{figure}
\begin{center}
 \subfigure{\includegraphics[width=0.49 \textwidth]{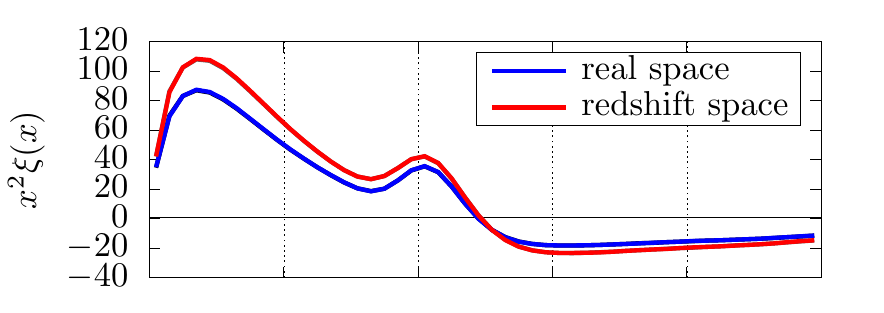}}
 \subfigure{\includegraphics[width=0.49 \textwidth]{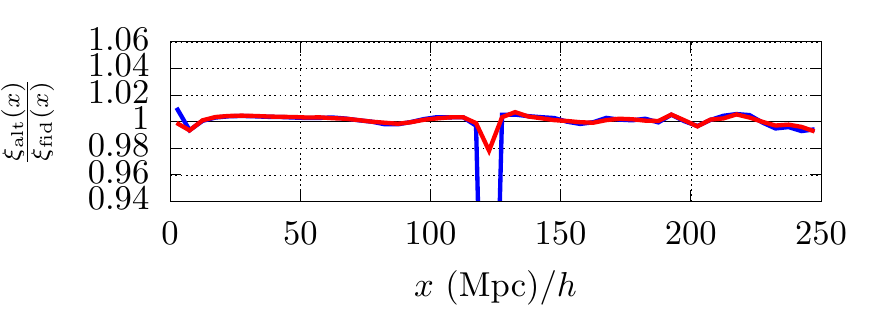}}
\end{center}
\caption{The upper panel: comparison between the correlation function of the set of 3000 EZmock boxes with different bispectrum (off by up to 20\%) and the one of the reference set (see Fig. \ref{fig:bk_offBK}). We show comparisons in both real space and redshift space. The references are shown with black and grey lines, but they are hardly visible since the second set has almost perfect agreement with the reference. The bottom panel: the ratios of the means of these two sets of EZmocks. We label the reference one with ``fid'' and the other one with ``alt''.}
\label{fig:cf_offBK}
\end{figure}

\begin{figure}
\begin{center}
 \subfigure{\includegraphics[width=0.49 \textwidth]{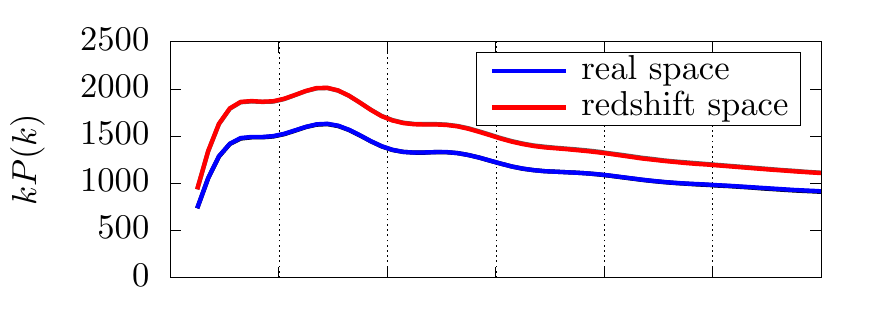}}
 \subfigure{\includegraphics[width=0.49 \textwidth]{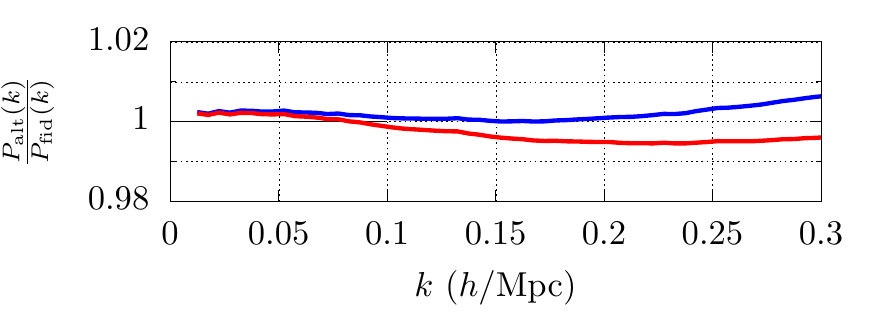}}
\end{center}
\caption{The upper panel: comparison between the power spectrum of the set of 3000 EZmock boxes with different bispectrum (off by 20\%) and the one of the reference set. We show comparisons in both real space and redshift space. The references are shown with black and grey lines, but they are hardly visible since the second set has almost perfect agreement with the reference. The bottom panel: the ratios of the means of these two sets of EZmocks. We label the reference one with ``fid'' and the other one with ``alt''.}
\label{fig:pk_offBK}
\end{figure}

\begin{figure}
\begin{center}
 \subfigure{\includegraphics[width=0.49 \textwidth]{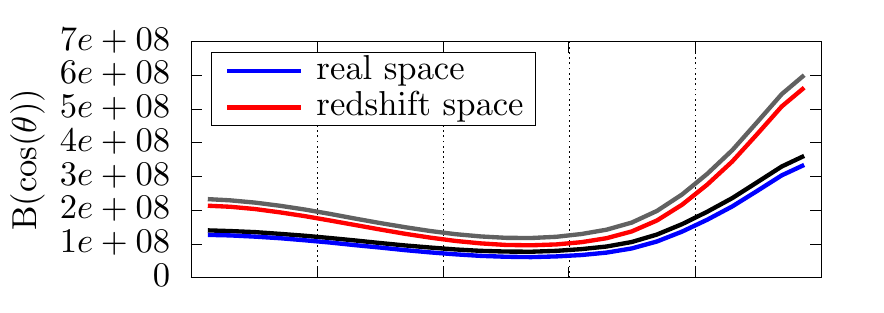}}
 \subfigure{\includegraphics[width=0.49 \textwidth]{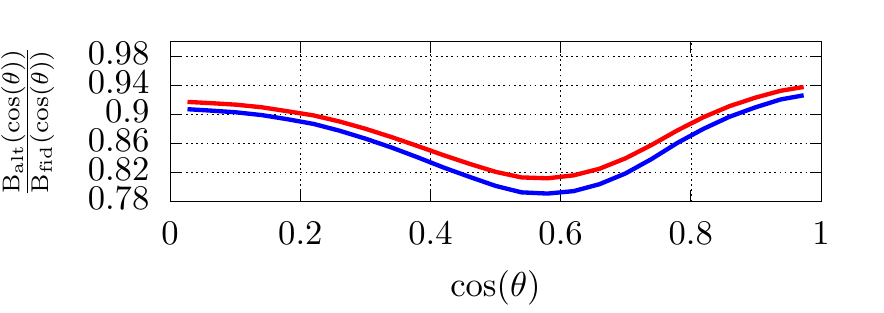}}
\end{center}
\caption{The upper panel: comparison between the bispectrum of the set of 3000 EZmock boxes with different bispectrum (off by 20\%) and the one of the reference set.  The configuration of the bispectrum is $\{k_1=0.1,k_2=0.2\}$. We show comparisons in both real space and redshift space. The bottom panel: the ratios of the means of these two sets of EZmocks. We label the reference one with ``fid'' and the other one with ``alt''.}
\label{fig:bk_offBK}
\end{figure}

The third set of 3000 EZmocks use the same input power spectrum as the reference one, but different bias parameters. We choose the  EZmock bias parameters of the third set to have the same 2-point but different 3-point clustering statistics as the reference set of EZmocks. Fig.\,\ref{fig:cf_offBK}, \ref{fig:pk_offBK}, and \ref{fig:bk_offBK} show the comparison of the correlation function, power spectrum, and bispectrum of these two sets of mocks. One can see that their agreement is within $1\%$ in power spectrum for $k<0.3$ and almost identical in most scales of the 2-point correlation function. The deviation of the bispectrum is up to 20\%. 

\section{results}
\label{sec:results}
\begin{figure}
\begin{center}
 \subfigure{\includegraphics[height=0.38 \textwidth,width=0.49 \textwidth]{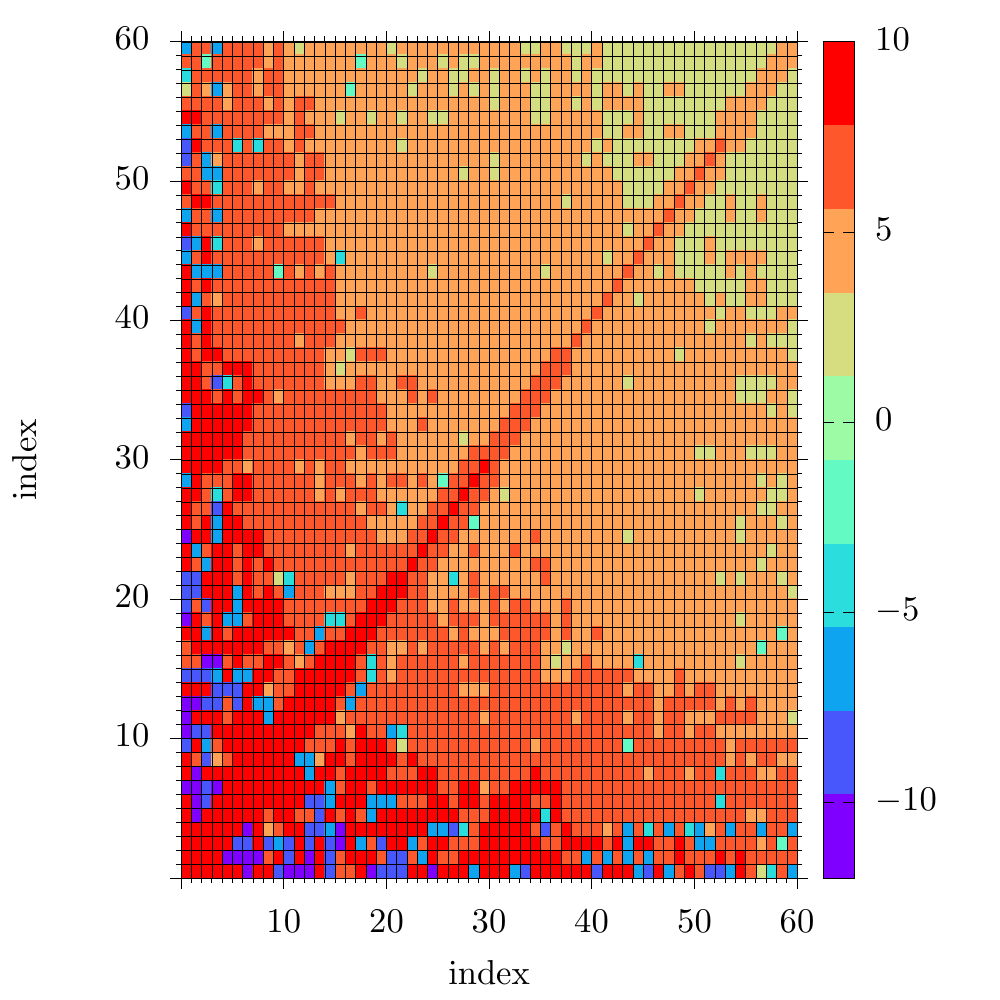}}
 \subfigure{\includegraphics[height=0.38 \textwidth,width=0.49 \textwidth]{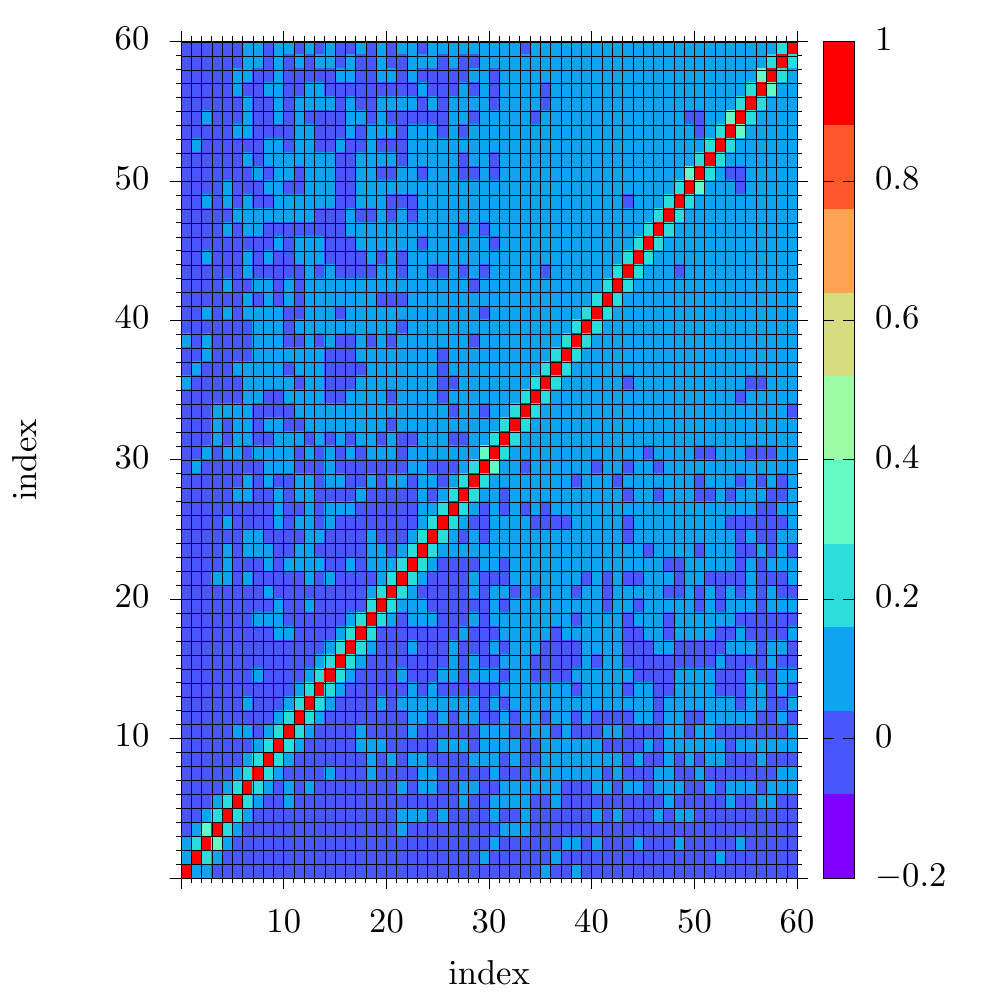}}
\end{center}
\caption{The covariance matrix (upper panel) and normalized covariance matrix (lower panel) of the power spectrum measured from the reference set of the EZmock boxes. The axes show the indices of the matrices. There are 60 bins within the range $0<k<0.3h$Mpc$^{-1}$ (bin size $=0.005h$Mpc$^{-1}$). The covariance matrix is rescaled by sign$(C_{ij})\mathrm{ln}(C_{ij}$sign$(C_{ij})+1)$. The normalized covariance matrix is in the orginal scale.}
\label{fig:2d_covar}
\end{figure}

We compute the covariance matrices from each set of 3000 boxes by
\begin{equation}
 C_{ij}=\frac{1}{\texttt{Num}-1}\sum^\texttt{Num}_{k=1}(\bar{X}_i-X_i^k)(\bar{X}_j-X_j^k),
\label{eq:covmat}
\end{equation}
where $\texttt{Num}$ is the number of the mock catalogues (i.e. 3000), $\bar{X}_m$ is the
mean of the $m^\mathrm{th}$ element of the data vector (composed of the bins of the correlation function, power spectrum, or bispectrum) from the mock catalogues, and
$X_m^k$ is the value in the $m^\mathrm{th}$ elements of the vector from the $k^\mathrm{th}$ mock
catalogue. 
Then, we compare their diagonal terms. We compute also the normalized covariance matrix by
\begin{equation}
 N_{ij}=\frac{C_{ij}}{C_{ii}^{1/2}C_{jj}^{1/2}},
\label{eq:covmat_nor}
\end{equation}
 and compare the first off-diagonal terms, $N_{i,i-1}$. 
 Fig. \ref{fig:2d_covar} shows the covariance matrix and normalized covariance matrix of the power spectrum measured from the reference set of the EZmock boxes. To simplify the problem, instead of comparing the full 2D covariance matrix, we compare only the diagonal terms and the first off-diagonal terms among different sets of simulations.

\subsection{Impact of $\sigma_8$ on the covariance matrix}
\label{sec:sigma8}

\begin{figure}
\begin{center}
 \subfigure{\includegraphics[width=0.49 \textwidth]{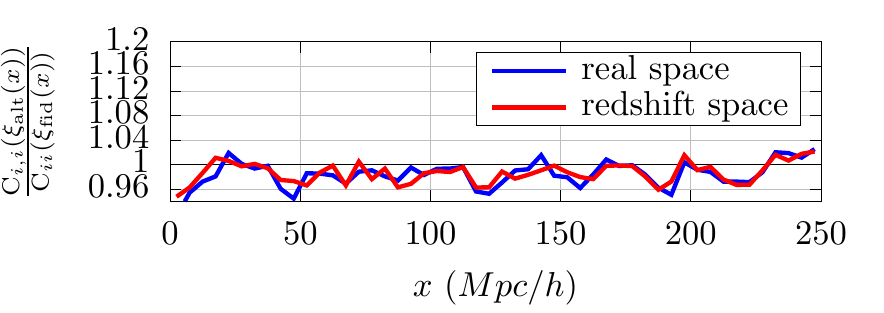}}
 \subfigure{\includegraphics[width=0.49 \textwidth]{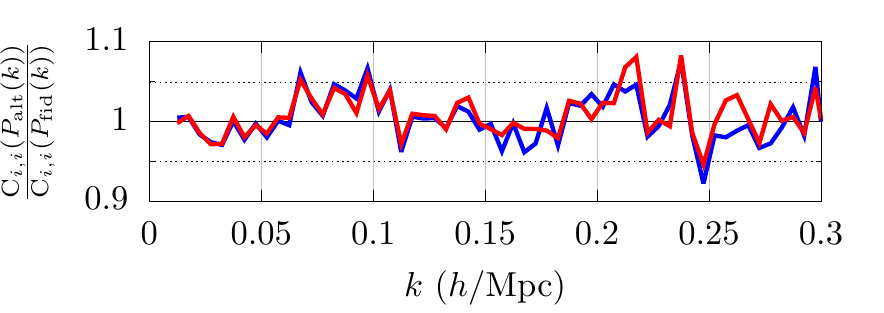}}
  \subfigure{\includegraphics[width=0.49 \textwidth]{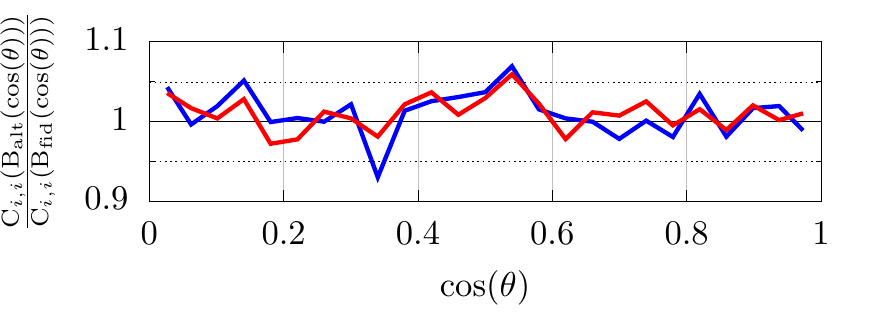}}
\end{center}
\caption{The ratios of the diagonal terms of the covariance matrices constructed based on the two sets of 3000 EZmock boxes. One set has 10\% smaller $\sigma_8$ than the reference set. We show both the cases in real space and redshift space. The upper panel shows the ratios of the correlation functions; middle panel shows the ones of the power spectrum; the bottom panel shows the ones of the bispectrum.}
\label{fig:diag_s8}
\end{figure}

\begin{figure}
\begin{center}
 \subfigure{\includegraphics[width=0.49 \textwidth]{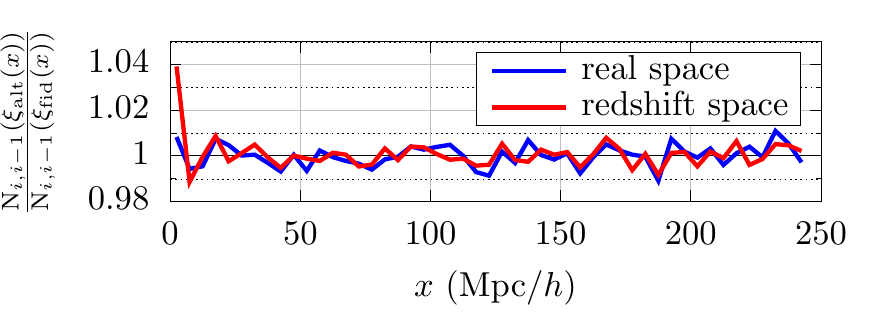}}
 \subfigure{\includegraphics[width=0.49 \textwidth]{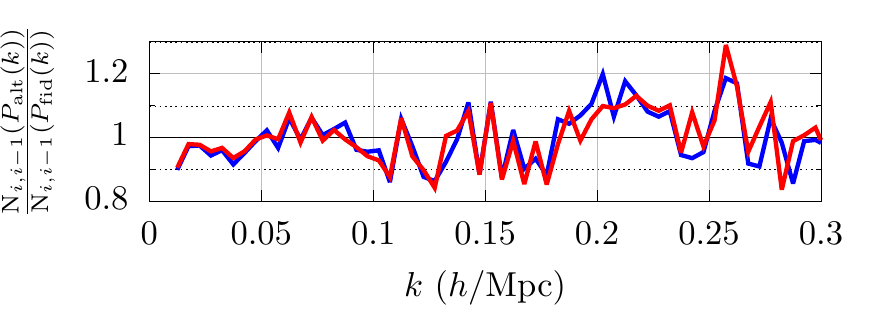}}
  \subfigure{\includegraphics[width=0.49 \textwidth]{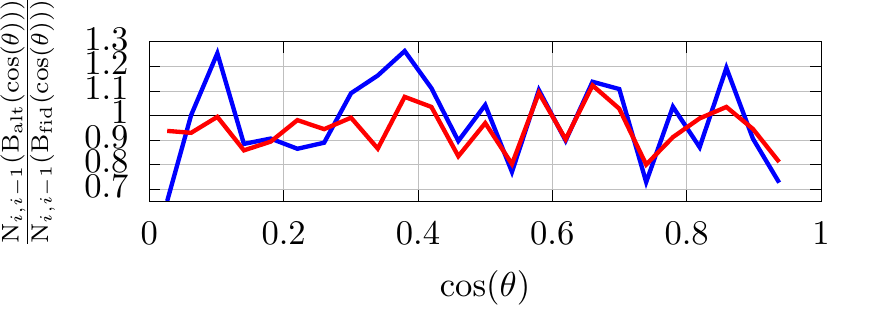}}
\end{center}
\caption{The ratios of the first off-diagonal terms of the normalized covariance matrices constructed based on the two sets of 3000 EZmock boxes. One set has 10\% smaller $\sigma_8$ than the reference set. We show both the cases in real space and redshift space. The upper panel shows the ratios of the correlation functions; middle panel shows the ones of the power spectrum; The bottom panel shows the ones of the bispectrum.}
\label{fig:corr_off1_s8}
\end{figure}

We compare the covariance matrix of the second set (lower $\sigma_8$ than the reference) with the reference. 
Fig.\,\ref{fig:diag_s8} and \ref{fig:corr_off1_s8} show the comparisons of their diagonal terms and the first off-diagonal terms (normalized), respectively.
Despite the 10\% difference in the input power spectrum and the slight difference in the mean of the clustering, we find that the covariance matrices agree with each other within 1 or 2\% for all the scales. 
Table \ref{table:s8} summarizes Fig. \ref{fig:diag_s8} and \ref{fig:corr_off1_s8} by showing the mean (subtracted by 1), standard deviation, and standard deviation of the mean (i.e. standard deviation divided by the square root of the number of bins) of the flat regions (with noise around a constant). One can see that all the means of the diagonal terms are within 1.5\% from zero. The first off-diagonal terms has a larger deviation, i.e. 5\%, which is not significant given larger uncertainty.

We conclude that, to construct a robust covariance matrix of a given observed clustering measurement, it is not critical to generate the mock catalogues with the true cosmology. Although we change only the input $\sigma_8$ value, not only the amplitude, but also the overall shape of the late-time clustering is actually different, e.g. see Fig. \ref{fig:pk_s8}. Thus, our conclusion should be able to be generalized to the cases using an input power spectrum with different shape (e.g., different matter fraction). This suggests that one can prepare the dark matter density fields for constructing covariance matrices without having concerns regarding the difference between the actually used cosmology and the true one.

\begin{table}
  \begin{tabular}{ | l | c | c | c | c | c | c | c |}     
    \hline    
  statistics    & ratio                                          & $\textrm{mean}-1$   & std. dev. & $\frac{\textrm{std.dev.}}{\sqrt{N}}$ \\ \hline  \hline
CF/real     & $C_{i,i}^{alt}/C_{i,i}^{\mathrm{fid}}  $  &  -0.0134 & 0.0179    &  0.0028  \\ \hline
CF/redshift & $C_{i,i}^{alt}/C_{i,i}^{\mathrm{fid}}  $  & -0.0136 & 0.0165    &  0.0026  \\ \hline
CF/real     & $N_{i,i-1}^{alt}/N_{i,i-1}^{\mathrm{fid}}  $  & -0.00020& 0.00454  &  0.00069 \\ \hline
CF/redshift & $N_{i,i-1}^{alt}/N_{i,i-1}^{\mathrm{fid}} $ & -0.00002 & 0.00367  &  0.00055 \\ \hline
\hline
PK/real     & $C_{i,i}^{alt}/C_{i,i}^{\mathrm{fid}}  $  & 0.0041 & 0.0305    &  0.0040 \\ \hline
PK/redshift & $C_{i,i}^{alt}/C_{i,i}^{\mathrm{fid}}  $  & 0.0086 & 0.0273    &  0.0036 \\ \hline
PK/real     & $N_{i,i-1}^{alt}/N_{i,i-1}^{\mathrm{fid}}  $  & 0.000  & 0.088     &  0.012  \\ \hline
PK/redshift & $N_{i,i-1}^{alt}/N_{i,i-1}^{\mathrm{fid}}  $  & 0.004  & 0.088     &  0.012  \\ \hline
\hline
BK/real     & $C_{i,i}^{alt}/C_{i,i}^{\mathrm{fid}}  $  & 0.0105       & 0.0280    & 0.0056 \\ \hline
BK/redshift & $C_{i,i}^{alt}/C_{i,i}^{\mathrm{fid}}  $  & 0.0106     & 0.0208    & 0.0042 \\ \hline
BK/real     & $N_{i,i-1}^{alt}/N_{i,i-1}^{\mathrm{fid}}  $  & -0.007 & 0.167     &  0.035 \\ \hline
BK/redshift & $N_{i,i-1}^{alt}/N_{i,i-1}^{\mathrm{fid}} $ & -0.051  & 0.096     &  0.020 \\ \hline
  \end{tabular}
  \caption{We summarize Fig. \ref{fig:diag_s8} and \ref{fig:corr_off1_s8} by computing the mean (subtracted by 1), standard deviation, and standard deviation of the mean (i.e. standard deviation divided by the square root of the number of bins) in the flat regions (with noise). Note that, for the correlation function (CF), we take the scale ranges of $x\in[50,250]\mathrm{Mpc}/h$ and $[25,250]\mathrm{Mpc}/h$ , containing $N=40$ and $44$ bins in the case of diagonal terms and the first off-diagonal terms, respectively;
for the power spectrum (PK), we take $k_i\in[0.01,0.3]h/\mathrm{Mpc}$ containing $N=58$ bins;
for the bispectrum (BK), we take $\mathrm{cos}(\theta)_i\in [0.0,1.0]$ and $[0.1,1.0]$, containing $N=25$ and $22$ bins in the case of diagonal terms and the first off-diagonal terms, respectively. One can see that all the means of the diagonal terms are within 1.5\%. The first off-diagonal terms have larger deviations, i.e. 5\%, which is not significant.}
\label{table:s8}
\end{table}


\subsection{Impact of the bispectrum on the covariance matrix}
\label{sec:bk}

\begin{figure}
\begin{center}
 \subfigure{\includegraphics[width=0.49 \textwidth]{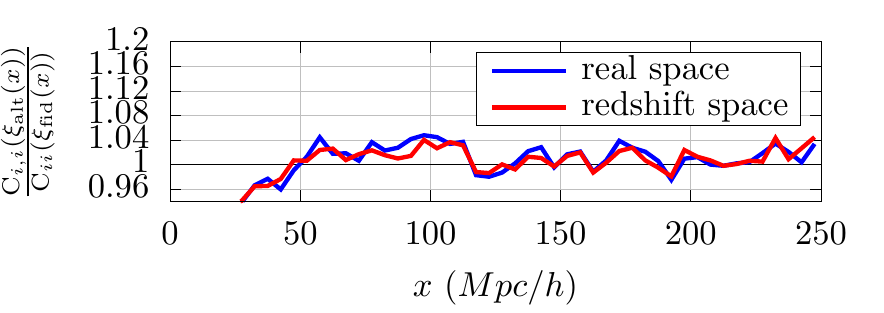}}
 \subfigure{\includegraphics[width=0.49 \textwidth]{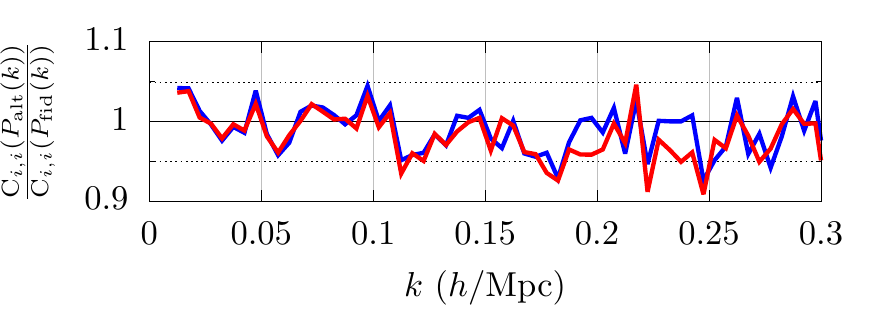}}
  \subfigure{\includegraphics[width=0.49 \textwidth]{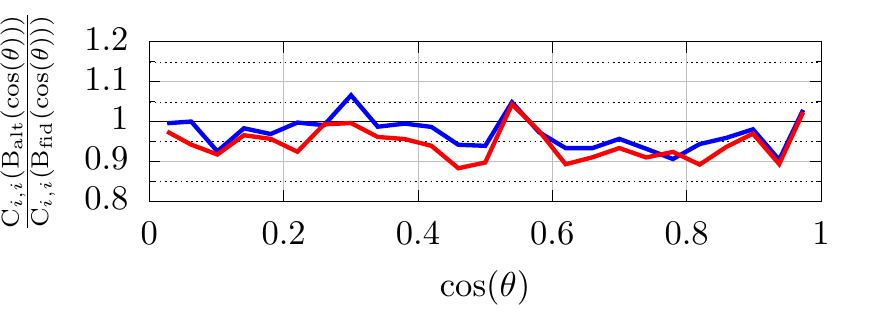}}
\end{center}
\caption{The ratios of the diagonal terms of the covariance matrices constructed based on two sets of 3000 EZmock boxes. One set has a different bispectrum (off by 20\%) from the reference set. We show the cases in both real space and redshift space. The upper panel shows the ratios of the correlation functions; the middle panel shows the ones of the power spectrum; the bottom panel shows the ones of the bispectrum.}
\label{fig:diag_offBK}
\end{figure}

\begin{figure}
\begin{center}
 \subfigure{\includegraphics[width=0.49 \textwidth]{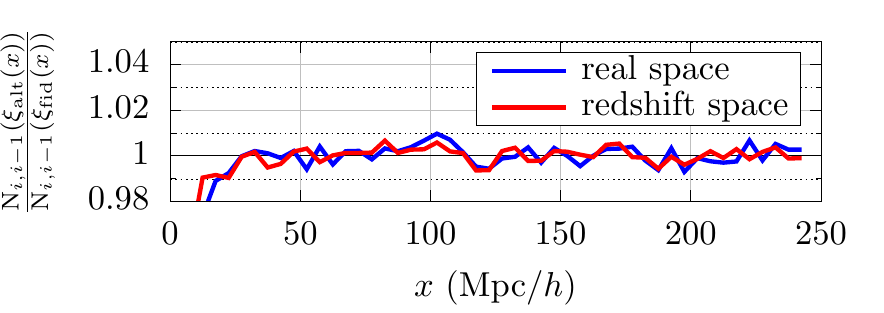}}
 \subfigure{\includegraphics[width=0.49 \textwidth]{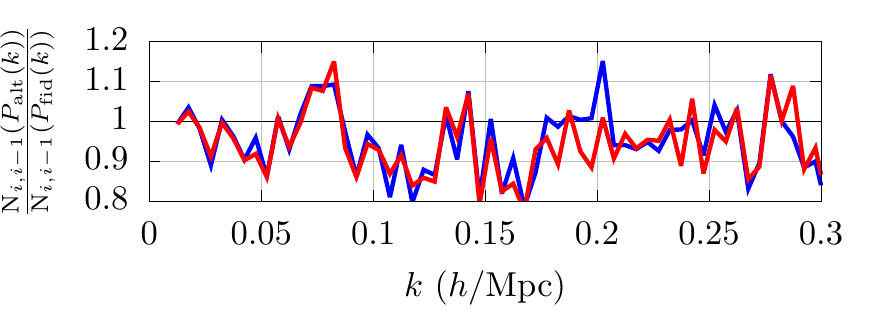}}
  \subfigure{\includegraphics[width=0.49 \textwidth]{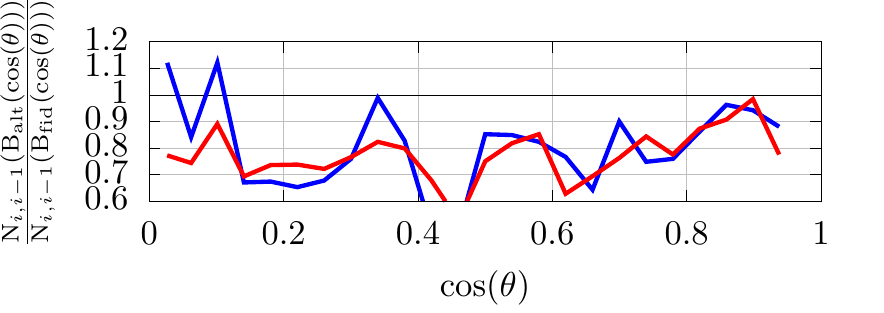}}
\end{center}
\caption{The ratios of the first off-diagonal terms of the normalized covariance matrices constructed based on two sets of 3000 EZmock boxes. One set has a different bispectrum (off by 20\%) from the reference set. We show the cases in both real space and redshift space. The upper panel shows the ratios of the correlation functions; middle panel shows the ones of the power spectrum; the bottom panel shows the ones of the bispectrum.}
\label{fig:corr_off1_offBK}
\end{figure}

We compare the covariance matrix of the third set (bispectrum different from the reference) with the reference. 
Fig.\,\ref{fig:diag_offBK} and \ref{fig:corr_off1_offBK} show the comparisons of their diagonal terms and the first off-diagonal terms, respectively.
While the agreement between the 2-point correlation functions is almost perfect, one can see that there is an obvious deviation between the covariance matrices at scales $<40h^{-1}$Mpc. 
Table \ref{table:bk} summarizes Fig. \ref{fig:diag_offBK} and \ref{fig:corr_off1_offBK} by computing the mean (subtracted by 1), standard deviation, and standard deviation of the mean (i.e. standard deviation divided by the square root of the number of bins) of the flat regions. One can see that the means from the diagonal terms of the 2-point statistics (i.e., correlation function and power spectrum) are within 2\% and the means from the first off-diagonal terms of the normalized covariance matrix are within 5\%. Thus, the covariance matrices of the 2-point clustering statistics at large scales are robust.
However, the errors on the covariance matrix of the bispectrum are much larger, which is expected.

Based on what we find, it is robust to extract cosmological constraints using 2-point correlation functions or power spectrum at large scales, even in the case that the mock catalogues do not reproduce accurately the 3-point clustering statistics from observed data (e.g. being off by 20\%). 
On the other hand, 3-point clustering can play an important role in determining the covariance matrices of the 2-point correlation functions at smaller scales, e.g., $<40h^{-1}$Mpc. Thus, one needs to be cautious when using the mock catalogues adopting the methodologies typically calibrated based on only 2-point clustering measurements, e.g. Halo occupation distribution (HOD; see \citealt{Seljak:2000gq,Peacock:2000qk,Berlind:2001xk}) and Subhalo abundance matching (SHAM; see \citealt{Vale:2004yt}). 
These methodologies rely on some assumptions, i.e. the galaxy-halo relation, which might not be accurate and result in the wrong 3-point clustering statistics, even if they reproduce reasonable 2-point clustering statistics.

In addition, the BAO reconstruction technique should require accurate 3-point and higher order clustering statistics. The mismatch in 3-point clustering statistics might lead to a mismatch in 2-point clustering statistics after applying the reconstruction methodology. In this case, the BAO measurement would be biased.

\begin{table}
  \begin{tabular}{ | l | c | c | c | c | c | c | c |}    
    \hline    
  statistics  & ratio                                         & $\textrm{mean} - 1$   & std. dev. & $\frac{\textrm{std.dev.}}{\sqrt{N}}$ \\ \hline    \hline
 CF/real        & $C_{i,i   }^{alt}/C_{i,i   }^{\mathrm{fid}}  $  &  0.0156 & 0.0189    &  0.0030  \\ \hline
 CF/redshift & $C_{i,i   }^{alt}/C_{i,i   }^{\mathrm{fid}}  $  &  0.0124 & 0.0159    &  0.0025  \\ \hline
 CF/real        & $N_{i,i-1 }^{alt}/N_{i,i-1 }^{\mathrm{fid}}  $  & 0.00062 & 0.00391  &  0.00059 \\ \hline
 CF/redshift & $N_{i,i-1 }^{alt}/N_{i,i-1 }^{\mathrm{fid}}  $  & 0.00038 & 0.00311  &  0.00047 \\ \hline
\hline
 PK/real        & $C_{i,i   }^{alt}/C_{i,i   }^{\mathrm{fid}}  $  & -0.0094 & 0.0293    &  0.0039 \\ \hline
 PK/redshift & $C_{i,i   }^{alt}/C_{i,i   }^{\mathrm{fid}}  $  & -0.0174  & 0.0298    &  0.0039 \\ \hline
 PK/real        & $N_{i,i-1 }^{alt}/N_{i,i-1 }^{\mathrm{fid}}  $  & -0.044   & 0.082     &  0.011  \\ \hline
 PK/redshift & $N_{i,i-1 }^{alt}/N_{i,i-1 }^{\mathrm{fid}}  $  & -0.052   & 0.081     &  0.011  \\ \hline
   \hline
 BK/real        & $C_{i,i   }^{alt}/C_{i,i   }^{\mathrm{fid}}  $  & -0.0288 & 0.0410    & 0.0082 \\ \hline
 BK/redshift & $C_{i,i   }^{alt}/C_{i,i   }^{\mathrm{fid}}  $  & -0.0553  & 0.0422    & 0.0084 \\ \hline
 BK/real        & $N_{i,i-1 }^{alt}/N_{i,i-1 }^{\mathrm{fid}}  $  & -0.211  & 0.155     &  0.033 \\ \hline
 BK/redshift & $N_{i,i-1 }^{alt}/N_{i,i-1 }^{\mathrm{fid}}  $  & -0.225 & 0.100     &  0.021 \\ \hline
  \end{tabular}
  \caption{We summarize Fig. \ref{fig:diag_offBK} and \ref{fig:corr_off1_offBK} by computing the mean (subtracted by 1), standard deviation, and standard deviation of the mean (i.e. standard deviation divided by square root of the number of bins) of the flat parts. Note that, for the correlation function, we take the scale ranges of $x\in[50,250]\mathrm{Mpc}/h$ and $[25,250]\mathrm{Mpc}/h$ , containing $N=40$ and $44$ bins in the case of diagonal terms and the first off-diagonal terms, respectively;
for the power spectrum, we take $k_i\in[0.01,0.3]h/\mathrm{Mpc}$ containing $N=58$ bins;
for the bispectrum, we take $\mathrm{cos}(\theta)_i\in [0.0,1.0]$ and $[0.1,1.0]$, containing $N=25$ and $22$ bins in the case of diagonal terms and the first off-diagonal terms, respectively. One can see that the means from the diagonal terms of the 2-point statistics (i.e., correlation function and power spectrum) are within 2\% and the means from the first off-diagonal terms of the normalized covariance matrix are within 5\%. The errors introduced by these deviations should be very small.
However, the errors on the covariance matrices of the bispectrum are much larger.
}
\label{table:bk}
\end{table}

\section{Conclusion and discussion}
\label{sec:conclusion}

In this work, we have tested the sensitivity of the covariance matrix to different factors, namely a different input power spectrum and a differently biased sample. By using the effective Zel'dovich approximation mock catalogues (EZmocks), which provide an efficient way to generate massive mock catalogues with accurate one-, two-, and three-point clustering statistics, we are able to minimize the discrepancy of the mean clustering measurements among the different sets of galaxy catalogues, so that one can compare the covariance matrices self-consistently.  
We have shown that the covariance matrix is insensitive to the input power spectrum, as long as the mock catalogues reproduce the observed clustering measurements, including one-, two-, and three-point statistics. In addition, with the same initial condition, we construct two sets of galaxy catalogues which have the same 2-point statistics (i.e. power spectrum and 2-point correlation function) but different bispectrum, and found that the disagreement in 3-point statistics introduces an obvious discrepancy in the covariance matrix of the 2-point correlation function at smaller scales, e.g., $r < 40 h^{-1}$Mpc. 
On the other hand, the covariance matrix can be still considered as robust at large scales, which suggests that it is not necessary to construct high precision mock catalogues reproducing the observed 3-point statistics when analysing the 2-point clustering statistics at large scales, e.g. measuring BAOs (Baryon Acoustic Oscillations) or RSDs (Redshift Space Distortions) with large-scale clustering measurements.
This is a good news for very large galaxy surveys, e.g. DESI, Euclid, LSST, and WFIRST, since one can minimize the effort and computational cost to construct reliable covariance matrices with efficient methodologies. 
On the other hand, in the studies using measurements from small scales, 
one should be cautious of the potential biases due to the mismatch in the 3-point statistics when using the mock catalogues calibrated based on only 2-point clustering measurements, e.g. HOD and SHAM.

In conclusion, an accurate estimation of galaxy bias, or an accurate cosmological parameter set is not compulsory to make precision cosmological analysis from galaxy clustering, as long as the 2 and 3-point statistics are accurately fitting observations, since then systematic deviations in both quantities compensate each other yielding unbiased covariance matrices.

\section{Acknowledgement}
We thank Francisco-Shu Kitaura, Volker M{\"u}ller, and Risa Wechsler for useful discussions. We thank Cheng Zhao for sharing the code for computing bispectra.
Main computation in this study has been done on the supercomputer JURECA at Julich Supercomputing Centre (project ID: HPO20).
We acknowledge PRACE for awarding us access to resource SuperMuc supercomputer based in Germany (project ID: 2010PA3442).



\label{lastpage}

{\small
\bibliography{covar_mat}
}

\end{document}